\documentstyle[aps]{revtex}

\begin{document} 
\draft

\title{Dynamics of Domains in Diluted Antiferromagnets}

\author{U.~Nowak, J.~Esser, and K.~D.~Usadel} 
\address{ 
Theoretische Tieftemperaturphysik\\ 
Gerhard-Mercator-Universit\"{a}t-Duisburg\\
47048 Duisburg/ Germany\\
e-mail: uli@thp.uni-duisburg.de 
} 

\date{\today}
\maketitle

\begin{abstract} 
  We investigate the dynamics of two-dimensional site-diluted Ising
  antiferromagnets. In an external magnetic field these highly
  disordered magnetic systems have a domain structure which consists
  of fractal domains with sizes on a broad range of length scales. We
  focus on the dynamics of these systems during the relaxation from a
  long-range ordered initial state to the disordered fractal-domain
  state after applying an external magnetic field. The equilibrium
  state with applied field consists of fractal domains with a size
  distribution which follows a power law with an exponential cut-off.
  The dynamics of the system can be understood as a growth process of
  this fractal-domain state in such a way that the equilibrium
  distribution of domains develops during time.  Following these ideas
  quantitatively we derive a simple description of the time
  dependence of the order parameter.  The agreement with simulations
  is excellent.
\end{abstract}


\section{introduction}

One starting point for the theoretical investigation of the behavior
of random magnets and related systems is to consider the system as
consisting of finite ordered regions, called clusters, domains or
droplets. In this article we want to consider the case of highly
disordered systems where the magnetic structure consists of fractal
domains with domain sizes on a broad range of length scales and we
will derive a simple theoretical description based on scaling ideas.
Nearly all of the assumptions we make and scaling relations we use
will be proven in detail by exact ground state calculations as well
as by Monte Carlo simulations.

The system we consider is the 2D diluted Ising antiferromagnet in an
external magnetic field (DAFF) which can be used to study
typical behavior of strongly disordered systems as there are
metastability, slow dynamics, and domain structures \cite{review}.
Related systems like random-field systems which are thought to
be in the same universality class like the DAFF \cite{fishman} or spin
glasses may behave similar.

The Hamiltonian of the DAFF in units of the next-neighbor coupling
constant $J$ reads
\begin{equation} 
\label{e.hamdaff} 
  H = \sum_{\langle i,j\rangle} \epsilon_i \epsilon_j \sigma_i
  \sigma_j -\sum_{i=1}^{N} B \epsilon_i \sigma_{i}
\end{equation}
where $\sigma_i = \pm 1$ denote Ising-spins and $\epsilon_i = 0,1$ a
quenched disorder. 

The phase diagram of the 2D DAFF consists of an antiferromagnetic low
temperature phase for magnetic field $B=0$ and a disordered phase for
all finite values of $B$. In contrast, in three dimensions there
exists a long-range ordered phase also for small magnetic fields
\cite{imbrie}. For higher fields and low temperatures the DAFF
develops a disordered domain state, both in two and three dimensions.
This domain state has many of the characteristics of a spin glass, as
for instance a remanent magnetisation and an irreversibility line
scaling like the deAlmeida-Thouless line \cite{nowak1}. For large
disorder the domains have been shown to have a fractal structure
\cite{nowak2} with a broad distribution of domain sizes and with
scaling laws strongly deviating from the usual Imry-Ma assumptions
\cite{imry} which are thought to be correct in the limit of small
disorder.

We focus on the dynamics of the 2D DAFF during relaxation from a
long-range ordered system to the fractal-domain state and describe
this process in terms of a certain growth process dominated by thermal
activation. In order to prove this approach in detail we use two
numerical techniques to investigate 2D systems with a size of $700
\times 700$ on a square lattice and a dilution of 30\%.

In Monte Carlo (MC) simulations \cite{binder} we use the standard
heat-bath algorithm. Due to the slow dynamics of the DAFF at low
temperatures it is extremely difficult to investigate equilibrium
properties using MC techniques.  Therefore, additionally we calculate
the highly non-trivial ground states of the DAFF exactly using
methods known from graph theory \cite{hartmann}. The Ising system is
mapped on a graph and the maximum flow through the graph is determined
by the Ford-Fulkerson algorithm \cite{swamy}. The combination of these
two methods, MC simulation and exact ground state (EGS) calculation,
allows a precise investigation of both dynamics and equilibrium
properties at finite temperatures and at $T=0$.

\section{Ground state properties}
The geometrical properties of domains in random magnets are important
as a starting point for theoretical considerations as well as for the
interpretation of experimental results.  In order to investigate the
structure of the domain state of a simulated DAFF we perform a cluster
analysis with a suitable adjusted Hoshen-Kopelman type algorithm
\cite{hoshen}.  This algorithm pieces the system into domains. A
domain is uniquely defined as a group of spins which are connected and
antiferromagnetically ordered. In this way, it is possible to compute
directly the volume $v$ (number of spins), surface $F$ (number of
unsatisfied bonds) and radius $R$ (root of the mean squared distance
of spins) of the domains formed (note that in two dimensions $v$ is
the area and $F$ the border of the domains).  Also one can compute the
corresponding energy-relevant quantities, like domain wall energy
$E_{w}$ (number of broken bonds) and volume-magnetization $M_{v}$ for
all domains.

The fractal properties of the domain state calculated by MC simulation
have been reported earlier \cite{nowak1,nowak2}. For high disorder we
discovered a fractal and interpenetrating structure of the domains
fulfilling scaling relations which are strongly deviating from the
assumptions of the Imry-Ma argument \cite{imry}. These earlier
findings are also correct for the EGS reported here for the first time
for the 2D DAFF. Fig.~\ref{mv} shows the volume-magnetisation $M_v(v)$,
i. e. the mean volume-magnetisation $M_v$ of domains with size $v$.
The scaling behavior is identical with that for the domains calculated
by MC simulation -- which are inevitably out of equilibrium. For 3D
systems this has been shown earlier \cite{hartmann}. As Fig.~\ref{mv}
shows the scaling relation $M_v(v) \sim v^{d_m}$ holds for large enough $v$.
All scaling relations of the DAFF obtained so far are summarized in
Table \ref{table}.

An important quantity is the distribution of domain sizes found in
the domain state of the 2D DAFF. In the next section we will use
this distribution function for a description of the dynamics of the
DAFF.  Fig.~\ref{nv} shows the number of domains $N$ for a given
volume $v$ from EGS calculations. Since the number of domains is a
strongly fluctuating quantity the data are averaged over intervals of
volumes $\Delta v$. The width of these intervals is increased
exponentially so that we have a constant distance between points on a
logarithmic scale. The data are well described by a power-law with an
exponential cut-off,
\begin{equation}
  N(v) = N_0 v^{-d_n} e^{-v/v_0}.
  \label{e.nv}
\end{equation}
The lines in Fig.~\ref{nv} are fits to Eq.~\ref{e.nv}.  The cut-off
parameter $v_0$ is field dependent. Although the fractal dimensions of
the domains of the DAFF are the same as those known from percolation
\cite{stauffer} the exponent $d_n \approx 1.5$ is between that for
critical percolation clusters ($d_n = 187/91$) and that for lattice
animals ($d_n = 1$). A possible small field dependence of this
exponent cannot be ruled out.  Also, the isolated clusters from the
percolation problem lead to the systematic deviation from the power
law in Fig.~\ref{nv} for very small volumes $v < 5$. Since there is no
infinite domain in the system we do not consider a streched
exponential cut-off in Eq.~\ref{e.nv} (the latter is known from the
percolation problem if there is an infinite cluster).  Summed up, the
domains of the DAFF have the structure of the lattice animals of the
percolation problem but with a different exponent $d_n$ in the
distribution function.

The distribution function Eq.~\ref{e.nv} is a central quantity of a
domain state since other quantities follow from this distribution. E.
g. the order parameter $M_s$ per spin can be rewritten as the sum over all
domain sizes $v$ of the corresponding order parameters times the
number of domains of that size,
\begin{equation}
  M_s = \int\limits_1^\infty \mbox{d}v ~n(v) M_s(v),
\end{equation}
where $n(v)$ is the distribution of domain sizes normalized in such a
way that $M_s$ is the staggered magnetisation per spin.

In the next section we will use the equations above as the starting
point for a description of the dynamics of the DAFF.

\section{Dynamics}
We focus on the dynamics of the two-dimensional DAFF. As was discussed
in the preceeding section the equilibrium state of the system in a
finite field consists of antiferromagnetically ordered domains with a
size distribution following Eq.~\ref{e.nv}. When we start the dynamics
with an antiferromagnetically long-range ordered system how does the
system evolve into this fractal-domain state?

Fig.~\ref{picture} shows a ground state configuration from EGS
calculations (bottom) and two configurations of the same system (i.~e.
with same configuration of vacancies) during the MC simulation. For
these pictures we restrict ourselves to a small system of size $100
\times 100$. We start the simulation with a long-range ordered state
(white) and switch on a field $B=1.5$ and a finite temperature
$T=0.4$.

The ground state of our system is essentially unique (see
Ref.~\cite{hartmann} for details). It is a domain pattern that is stored in
the system through the vacancy configuration and the field and it is
the equilibrium state of the system. How is this domain pattern
reconstructed in our Monte Carlo simulation?

As Fig.~\ref{picture} suggests, during the relaxation from the
long-range ordered initial state to the fractal-domain state the
dynamics of the system can be understood as a kind of growth process.
For shorter times only small domains arise.  Interestingly, these small
domains are located at those places where later the larger domains
will be. At later times, {\it also} larger domains occur and the size
of the largest domains increases in time.  The pictures lead us to the
following description of the growth process: the equilibrium
distribution of domains develops starting with the smaller domains and
increasing the maximum domain size in time.

In order to investigate this process quantitatively we simulate the
size distribution of domains during the relaxation process.
Fig.~\ref{nv_dyn} shows the equilibrium distribution from EGS
calculations and distributions of domains that are reversed after five
different times of MC simulation. For a given time $t$ the system has
obviously a distribution that is equal to the equilibrium distribution
$n(v)$ corresponding to Eq.\ref{e.nv} up to the largest domain size
$v_m(t)$ that has developed at that time by thermal activation. Note
that this time dependent cut-off in the distribution function is very
abrupt since Fig.~\ref{nv_dyn} shows data over a range of eight
decades.  These considerations lead to the following description of
the time development of the order parameter $M_s$.  We start with a
long-range ordered system, i.  e. $M_s(t=0) = M_{si} \approx 1$.
\footnote{Note that $\lim_{t \to 0} M_s < 1$ since due to the dilution
  there are spins or clusters of spins that are not or only weakly
  connected to the infinite cluster. These clusters may have finite
  magnetisations which follow the external field immediately since
  there are no relevant energy barriers.} After switching on the field
the staggered magnetisation decreases to $M_s(t \rightarrow \infty) =
M_{s \infty} \approx 0$ (except of finite-size effects) due to the
growth of domains of the other phase:
\begin{equation}
  M_s(t) = M_{si} + 2 \int\limits_1^{v_m(t)} \mbox{d}v ~ n(v) M_s(v)
\end{equation}

Since a domain is defined as a group of spins which are
antiferromagnetically ordered it is per definition $M_s(v) = -v$ for
the reversed domains so that
\begin{eqnarray}
  \label{e.gamma}
  M_s(t) & = & M_{si} - 2 \int\limits_1^{v_m(t)} \mbox{d}v ~ n_0
               v^{1-d_n} e^{-v/v_0} \nonumber\\
         & = & M_{s \infty} + 2 n_0 \int\limits_{v_m(t)}^\infty
               \mbox{d}v ~ v^{1-d_n} e^{-v/v_0} \nonumber\\
         & \equiv & M_{s \infty} + \gamma(v_m(t))
\end{eqnarray}
where the last line defines the function $\gamma(v_m(t))$.
$n_0$ is defined by the normalisation condition
\begin{equation}
  2 \int\limits_1^{\infty} \mbox{d}v \ n_0 v^{1-d_n} e^{-v/v_0} + M_{s
    \infty} = M_{si}.
\end{equation}
According to the preceeding section the exponent is $d_n = 1.5$ and
the field dependent cut-off parameter can be obtained from EGS
calculations. Thus $\gamma(v_m(t))$ is well defined without
any unknown parameters.  Since the integration cannot be done exactly
it will be calculated numerically.  Furthermore, the integral
$\gamma(v_m(t))$ is related to the so-called incomplete Gamma-function
\begin{equation}
  \Gamma(\alpha,x) \equiv \int\limits_0^x \mbox{d}y ~ y^{\alpha-1}
  e^{-y}
\end{equation}
the properties of which we will discuss later.

It is highly plausible that the largest existing domain size $v_m(t)$
for a time $t$ is connected to a thermal activation energy $\Delta
E(t) = T \ln (t/\tau) $ according to
\begin{equation}
  \label{e.ar}
  \Delta E(t) = E_b v_m^{d_b}(t)
\end{equation}
where $d_b$ is an exponent connecting the size of domains with a 
free-energy barrier $\Delta E$ which has to be overcome during the
build-up of a domain of size $v_m$. This assumption of thermal
activation gives the time dependence of $M_s$ in Eq.\ref{e.gamma}.

In order to prove the validity of our approach we use MC simulation
data for the decay of the order parameter. In Fig.~\ref{gamma} we show
data for three different values of the magnetic field $B=1.5, 2.0,
2.5$. The way we evaluate the simulation data is the following:

i) First of all, due to the equations above data for the relaxation
for different temperatures should collapse in a scaling plot $M_s$
versus any function $f(T \ln (t/\tau))$ where $\tau$ should be a
microscopic time scale for the growth of the smallest domains. In
Fig.~\ref{gamma} the data for four ($B = 1.5$) respectively three ($B
= 2.0, 2.5$) different temperatures and a time window ranging from 95
MCS to 200000 MCS are seen to collapse very well using $\tau = 3$ MCS.
We do not consider time scales much shorter than 100 MCS because then
thermal fluctuations cannot be neglected.

ii) Finding the correct barrier exponent $d_b$ and the prefactor $E_b$
we can determine $v_m(t) = ((T/E_b) \ln (t/\tau))^{1/d_b}$ and
following Eq.~\ref{e.gamma} the data for $M_s(t)-M_{s \infty}$ should
be described by $\gamma(v_m(t))$. As Fig.~\ref{gamma} shows this works
perfectly for the complete range of the scaling variable $v$ and for
three different fields using $d_b = 0.25$. Note that the reason for
the field dependence of $\gamma(v_m(t))$ is the field dependence of
the cut-off parameter $v_0$ which has been obtained by EGS
calculations (see Fig.~\ref{nv}).

The data for the two higher fields are less accurate since in this
case the dynamics is faster and has a shorter observation time.
Additionally for the higher fields we restricted ourselves to smaller
lattices ($400 \times 400$ instead of $700 \times 700$ for $B = 1.5$)
since the EGS calculations are very time consuming. However, no field
dependence of the barrier exponent can be seen. The prefactor $E_b$ is
slightly field dependent ($E_b = 0.60 (B=1.5), 0.74 (B = 2.0), 0.67 (B
= 2.5))$. Note that since the barrier exponent is very small we are
close to the case $\Delta E(t) \sim \ln(v_m(t))$ which is the limit
$d_b \to 0$ in Eq.~\ref{e.ar}.

Of course, other quantities can be described in a similar manner. E.
g. for large enough volumes $v$ (see Fig.~\ref{mv}) the magnetisation
$M$ can be written correspondingly as
\begin{equation}
  \label{e.mt}
M(t) = M_i + 2 \int\limits_1^{v_m(t)} \mbox{d}v ~ n_0 M_0
           v^{d_m-d_n} e^{-v/v_0}
\end{equation}
The magnetisation $M$ is a quantity that can be measured
experimentally. Its precise behavior has to be obtained from a
numerical integration of Eq.~\ref{e.mt}. Two for the description of
experimental data relevant limits can however be obtained
asymptotically as limits of the Gamma-function.

The first limit is that of short times and small temperatures where
the exponential cut-off is irrelevant which means $((T/E_b)
\ln(t/\tau))^{1/d_b} = v \ll v_0$ and consequently $e^{-v/v_0} \approx
1$ resulting in
\begin{eqnarray}
  M(t) & \approx & M_i + 2 n_0 M_0 (v(t))^{d_m-d_n+1} \nonumber \\
       & =       & M_i + 2 n_0 M_0 (\frac{T}{E_b} \ln
                   (t/\tau))^{\frac{d_m-d_n+1}{d_b}}.
\end{eqnarray}
This is a logarithmic law with a temperature-independent exponent
obtained from the fractality and the distribution of the equilibrium
domains and from the free-energy barriers that have to be overcome to
build up these domains. Similar logarithmic laws are usually used to
describe experimental data of the dynamics of the DAFF (for a review
of the experimental work see \cite{kleemann}).  Note that $M_i$ is not
zero but has a small finite value for the same reason that has been
explained earlier -- see the discussion of $M_{si}$.

The second limit is that close to equilibrium, i. e. for long times and
higher temperatures. Here we can use an asymptotic form of the
Gamma-function
\[
\lim_{x \to \infty} \Gamma(\alpha,x) \sim x^{\alpha-1} e^{-x}.
\]
The restriction to the most relevant term -- the e-function -- results in 
\begin{eqnarray}
  M(t) & \sim & e^{-(\frac{T}{E_b} \ln
                (\frac{t}{\tau}))^{1/d_b}/v_0} \label{e.gp}\\
       & \sim & t^{- T/(E_b v_0)} \quad \mbox{for} \quad d_b = 1. \label{e.pl}
\end{eqnarray}
The generalized power law Eq.~\ref{e.gp} has been used to fit
experimental data as well as data from MC simulations for the remanent
magnetisation of the 3D DAFF \cite{han,han2}. For $d_b=1$ it simplifies to
a simple power law with the exponent proportional to the temperature.
This law has also been discussed as a description of the remanent
magnetisation of the 3D DAFF \cite{nowak3} and it is typical for the
decay of the remanent magnetisation of spin glasses (see \cite{rieger}
for a review).

\section{conclusions}
We derived a simple description for the dynamics of the 2D DAFF during
the relaxation from long-range order to a fractal-domain state.  The
basic fundaments of this description are the fractality and the
distribution of the domains the domain state of the system consists of
in the limit of high disorder. This is the main extension to earlier
theories for the case of weak disorder based on scaling assumptions
for domains which are compact and non-fractal with for a certain time
scale unique relevant length scale -- the domain radius (see e. g.
\cite{imry,villain} for scaling theories of random-field models and
\cite{villain,natter} for a description of domain dynamics in terms of
domain wall motion).

As a result of our approach which is based on measured quantities of
the domain state the dynamics can be described with the assumption of
thermal activation analytically and is given by an incomplete
Gamma-function.  As one advantage this function connects naturally two
for disordered systems often observed laws as limiting cases --- the
logarithmic law in the limit of short times and low temperatures and
the (generalized) power law in the limit of late times and higher
temperatures. Apart from that our approach illustrates that from the
fact that a system has a distribution of domain sizes instead of one
dominating size follows that the mean domain size of that distribution
can grow following a power law (Eq.~\ref{e.pl}) even if the largest
domain in the system is build by thermal activation, i. e. within a
logarithmic time scale (see e. g. \cite{rieger} for a summary of this
apparent contradiction in spin glasses).

The agreement with simulations is excellent. However, so far we are
not aware of any direct experimental observation of this kind of
dynamics in 2D DAFF although as mentioned above similar laws are
often observed in the dynamics of disordered magnets and we assume
that a similar description might also work for other strongly
disordered systems like random-field systems or spin glasses.

\acknowledgements The authors wish to thank M. Staats for technical
help. The work was supported by the Deutsche Forschungsgemeinschaft
through Sonderforschungsbereich 166.

\begin{figure}
  \caption{Magnetisation $M_v$ versus volume $v$ of the domains of the
    fractal-domain state of the 2D DAFF. Comparison of MC simulations
    (avarage over 50 systems of size $199 \times 198$) and EGS
    calculations (avarage over 10 systems of size $400 \times 400$).
    The latter curve is shifted by a factor 2.  $B=1.5$. All
    quantities are dimensionless.}
  \label{mv}
\end{figure}

\begin{figure}
  \caption{$N$ versus $v$ from EGS calculations for different fields $B$.
    Apart from the average over 10 systems of size $400 \times 400$
    the data are additionally averaged in such a way that they show
    the relative number of domains having a volume within a certain
    intervall $\Delta v$. The lines are fits to Eq.~\ref{e.nv}.}
  \label{nv}
\end{figure}

\begin{figure}
  \caption{EGS of a $100\times 100$ system (bottom) and two
    configurations of the system during the MC simulation of the
    relaxation process after roughly 500 MCS (top) and 130000 MCS (center).
    $T=0.4$, $B=1.5$. The two antiferromagnetic phases are represented in
    black and white, the vacancies are grey.}
  \label{picture}
\end{figure}

\begin{figure}
  \caption{$N$ versus $v$ during MC simulation of the relaxation
    process and for the EGS of the same $700 \times 700$ system (see
    Fig.~\ref{nv} for explanation). $T=0.4$, $B=1.5$.}
  \label{nv_dyn}
\end{figure}

\begin{figure}
  \caption{Scaling plot for the order parameter $\Delta M_s =
    M_s(t)-M_{s\infty}$. Data are shown for three different fields, 
    $B = 1.5, 2.0, 2.5$ (from above) and
    four ($B=1.5$) respectively three ($B=2.0, 2.5$) different
    temperatures.  The time ranges from 95 MCS to 200000 MCS. The
    solid lines are the theoretical curves following Eq.~\ref{e.gamma}
    with $v_0 = 9000 (B=1.5), 1000 (B=2.0), 400 (B=2.5)$.}
  \label{gamma}
\end{figure}

\narrowtext
\begin{table}
  \caption{Scaling relations for the DAFF}
  \begin{tabular}{c||c|c|c|c|c}
    \renewcommand\arraystretch{1.}
    $D = 3$ & $v \sim R^{2.0}$  & $F \sim v$ & $M_v \sim v$ & $E_w
    \sim F$ & - \\ \hline
    $D = 2$ & $v \sim R^{1.56}$ & $F \sim v$ & $M_v \sim v$ &  $E_w
    \sim F$ & $n \sim v^{-1.5} \exp(-v/v_0)$
    \label{table}
  \end{tabular} 
\end{table}

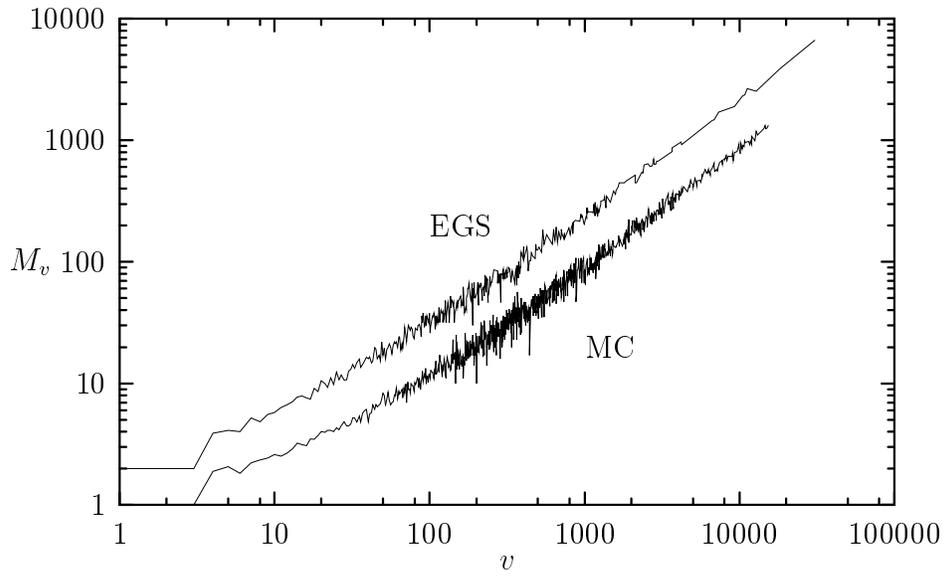
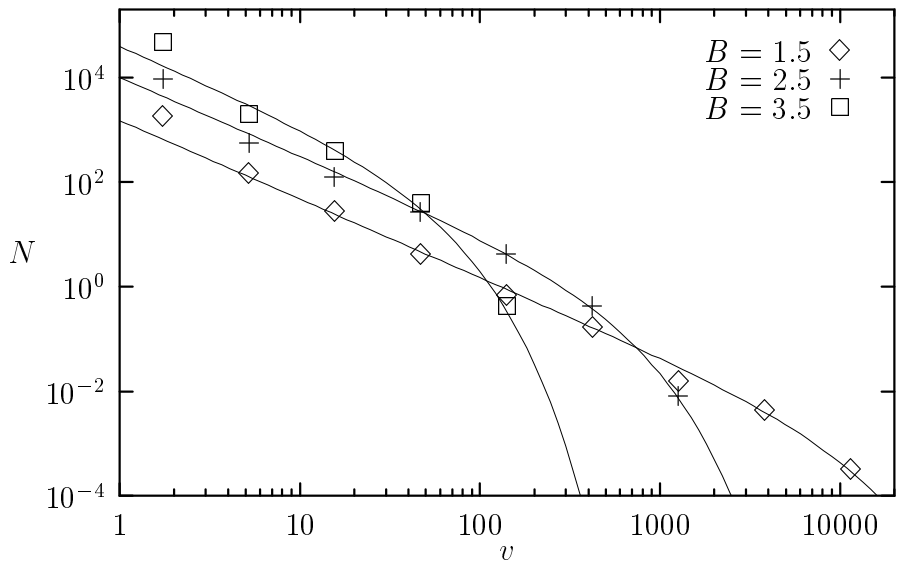
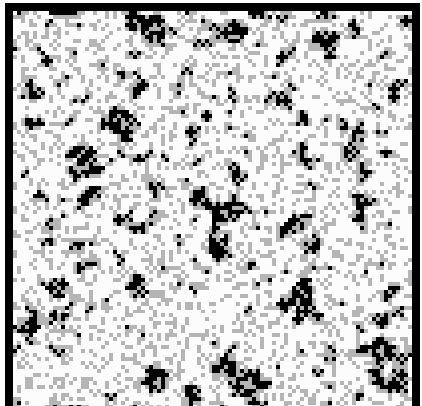
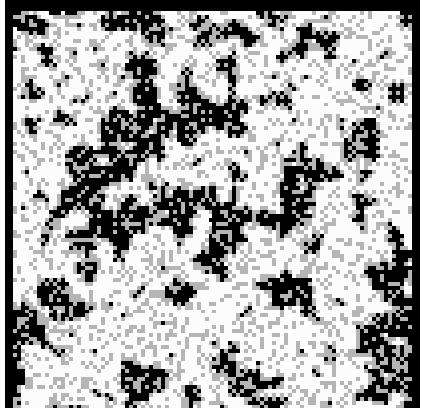
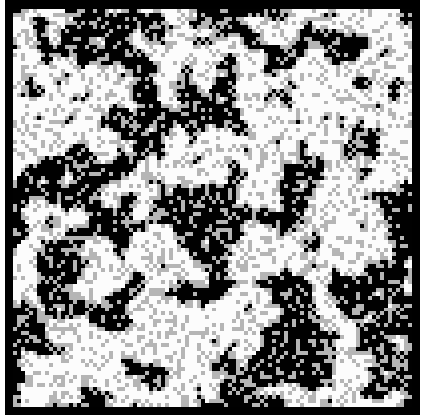
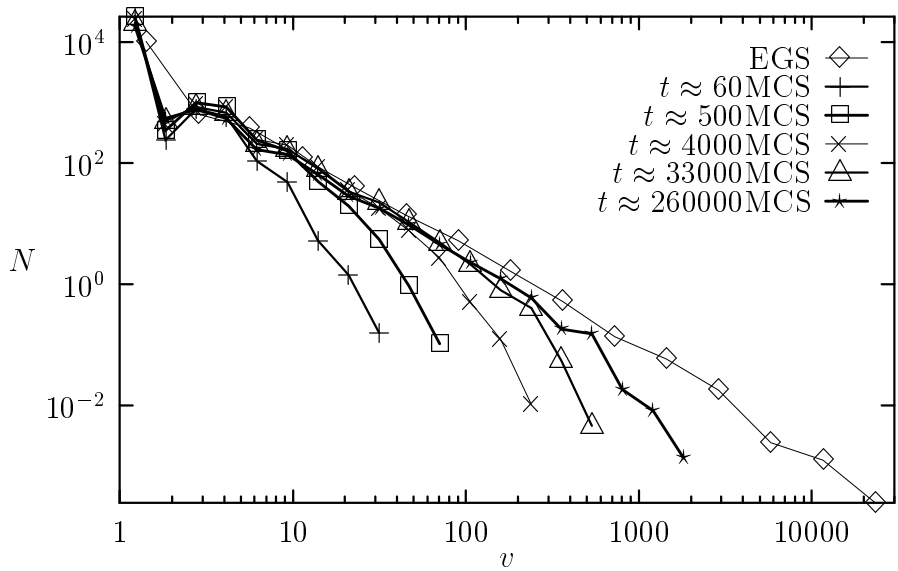
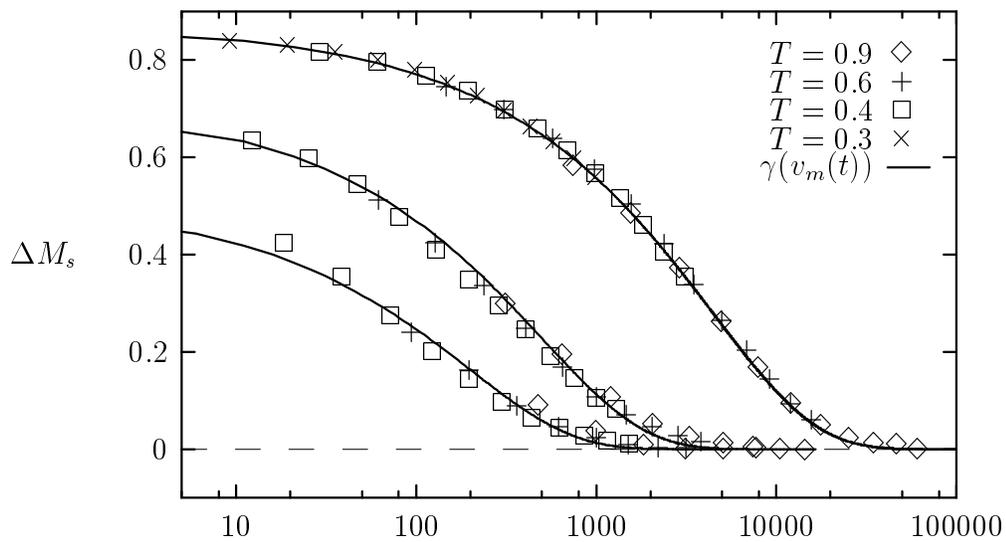
\end{document}